\RequirePackage[final]{graphicx}
\documentclass[aps, superscriptaddress,pra,twocolumn,footinbib]{revtex4-1}
\usepackage[utf8]{inputenc}
\usepackage{amsmath}
\usepackage{amsfonts}
\usepackage{mathtools}
\usepackage{graphicx}
\usepackage{cleveref}
\usepackage{fixme}
\usepackage{color}

\renewcommand{\k}[0]{\mathbf{k}}
\newcommand{\s}[0]{\sigma}

\newcommand{\bs}[0]{\bar{\sigma}}
\renewcommand{\d}[0]{\dagger}
\newcommand{\Q}[0]{\mathbf{Q}}
\newcommand{\q}[0]{\mathbf{q}}
\renewcommand{\eqref}[1]{(\ref{#1})}
\def\nn{\nonumber}

\begin{document}

\title{Stabilising Fulde-Ferrel-Larkin-Ovchinnikov Superfluidity with long-range Interactions in a mixed dimensional  Bose-Fermi System}
\date{February 2018}
%\author{J.~M. \surname{Midtgaard}}
\author{Jonatan Melk\ae r \surname{Midtgaard}}
\affiliation{Department of Physics and Astronomy, Aarhus University, DK-8000 Aarhus C,  Denmark}
\author{Georg M.\ \surname{Bruun}}
\affiliation{Department of Physics and Astronomy, Aarhus University, DK-8000 Aarhus C,  Denmark}

\begin{abstract}
We analyse the stability of inhomogeneous superfluid phases in  a system consisting of identical fermions confined in two layers that 
are immersed in a Bose-Einstein condensate (BEC).  The fermions in the two layers interact via an induced interaction mediated by the 
BEC, which gives rise to pairing. We present zero temperature phase diagrams varying 
the chemical potential difference between the two layers and the range of the induced interaction, and show that there is a  large region where an inhomogeneous 
superfluid phase is the ground state. This  region grows with increasing range of the induced interaction and it can be much larger than 
for a corresponding system with a short range interaction. The range of the interaction is controlled by the healing length of the BEC, 
which makes the present system a powerful tunable platform to stabilise  inhomogeneous superfluid phases. 
We furthermore analyse the melting of the superfluid phases in the layers via phase 
 fluctuations as described by the Berezinskii-Kosterlitz-Thouless mechanism and show that the normal, homogeneous and inhomogeneous 
 superfluid phases meet in a tricritical point. The superfluid density of the inhomogeneous superfluid phase is reduced by 
 inherent  gapless excitations, and we demonstrate that this leads to a significant suppression of the critical temperature as compared to the 
 homogeneous superfluid phase. 
\end{abstract}

\maketitle

%\underline{Hovedresultater:}
%\begin{itemize}
%\item Long range interactions good for FFLO. Much larger region in phase diagram than for short range. 
%\item FFLO + BKT. Get tri-critical point.
%\item Anisotropic superfluid density. 
%\item Can tune interaction so that maximum $T_c\sim T_F/8$ is obtained for BCS? Gives high $T_c$ for FF phase as well. 
%\end{itemize}
%
%\underline{Figurer:}
%
%\begin{enumerate}
%\item Setup cartoon. 
%\item Phase diagrams $(k_F\xi,h)$ for two different $g$'s. 
%\item $\Delta(Q,k_F)$ and $\Delta(0,k_F)$ plot versus $k_F\xi$.
%\item Phase diagram $(k_Fd,h)$.
%\item $n_\parallel$ and $n_\perp$ vs.\ $T/T_c$ plots. Cartoon expressing what they do physically. 
%\item Phase diagram $(T/T_c,h)$. Tri-critical point. 
%\end{enumerate}

\section{Introduction}
 The  interplay between population imbalance and superfluid pairing has been subject to intense study ever since 
 Fulde and Ferrell (FF) as well as Larkin and Ovchinnikov (LO) predicted that they can co-exist~\cite{Fulde1964,Larkin1965}. In condensed matter systems, 
 an external magnetic field leads to a population imbalance between the two electron  spin projections, which in general is at odds with superconductivity. 
  FFLO however realised that the superconductor can accommodate some population imbalance at the price of 
 giving the Cooper pairs a non-zero center-of-mass (COM) momentum, thereby forming a spatially inhomogeneous but periodic order parameter with no vortices.
 The fate of superfluid pairing in the presence of population imbalance  is a fundamental  question relevant for many systems in nature including 
 cold atoms~\cite{Chevy2010,Radzihovsky2010,Kinnunen2018},   superconductors~\cite{Matsuda2007}, and quark matter~\cite{Casalbuoni2004,Alford2008,Anglani2014}. 
Nevertheless, an unambiguous observation of a FFLO phase is still 
lacking. A major problem for electronic superconductors  is that orbital effects due to the magnetic field 
lead to the formation of vortices and eventually destroy pairing before any FFLO physics can be observed. 
 One strategy to avoid this problem is to explore low dimensional systems, where orbital effects are suppressed due to the confinement. 
  Indeed,  results consistent with  a FFLO phase have been reported for quasi-2D organic and heavy fermion 
  superconductors~\cite{Beyer2013,Mayaffre2014,Bianchi2003}. Theoretical studies have furthermore concluded that 
the FFLO phase is favored in 2D as compared to 3D~\cite{Shimahara1994,Burkhardt1994,Pieri2007}. 

Quantum degenerate atomic gases are well suited to investigate FFLO physics, because they do not suffer from orbital effects as the atoms are neutral. In addition, 
 it is relatively straightforward to make low dimensional systems, and signatures of FFLO physics has indeed been observed in a one-dimensional (1D) atomic Fermi gas~\cite{Liao2010}.
  There has been a number of investigations of the
FFLO phase for 2D atomic gases with a short range interaction~\cite{Gukelberger2016,Baarsma2016,Conduit2008,Radzihovsky2009,Wolak2012,Shaoyu2014,Sheehy2015}.
Recently, it was argued that long range  interactions further increases the region of stability of the FFLO phase for a 2D gas of dipolar atoms as compared to a short range 
interaction~\cite{Lee2017}.

Here, we investigate how to stabilise FFLO superfluidity using a mixed dimensional system with a tuneable range interaction. The system consists 
of fermions confined in two layers immersed in a BEC. 
The induced interaction between the layers mediated by the BEC gives rise to pairing, and we analyse the stability of the corresponding 
superfluid phases. We show that the zero temperature phase diagram as a function of the chemical potential difference in the two layers and the 
range of the interaction has a large region, where an inhomogeneous superfluid phase is the ground state. This region grows with increasing range of the interaction and becomes 
much wider than for a zero range interaction. The interaction range can be tuned by varying the healing length of the BEC. We furthermore investigate the melting of the 2D superfluid phases via the  Berezinskii-Kosterlitz-Thouless mechanism. The normal phase, the homogeneous and inhomogeneous 
superfluid phases are shown to meet in a tricritical point in the phase diagram, which determines the maximum critical temperature of the inhomogeneous superfluid. 
This maximum temperature is however significantly suppressed compared to the homogeneous superfluid phase, due to inherent gapless excitations, which decrease the superfluid density.

%%%%%%%%%%%%%%%%%%%%%%%%%%%%%%%%%%%%%%%%%%%%%%%%%%%%%%%%%%
\begin{figure}[htb]
\centering
\includegraphics[width=\columnwidth]{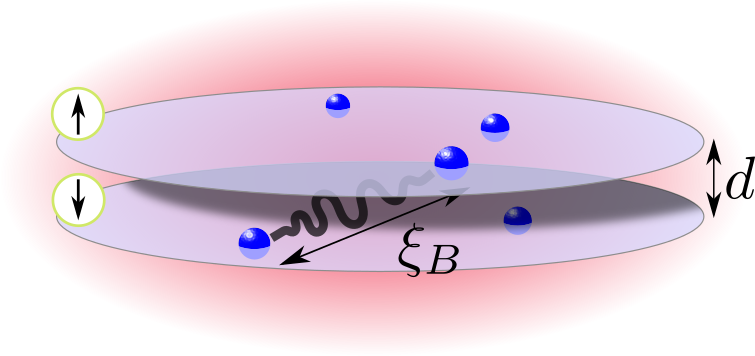}
\caption{(Color online) A sketch of the considered system. The  fermions (blue) are confined in two layers immersed in a three-dimensional BEC (red), with a layer distance $d$. The BEC mediates an interaction  between fermions in the two different layers of the Yukawa form, 
with a range determined by the BEC healing length $\xi_B$. }
\label{fig:system}
\end{figure}
%%%%%%%%%%%%%%%%%%%%%%%%%%%%%%%%%%%%%%%%%%%%%%%%%%%%%%%%%%
\section{The Bilayer System}
We consider the system illustrated in Fig.~\ref{fig:system}. Two layers contain fermions of a single species with mass $m$, and they are 
 immersed in a 3D weakly interacting BEC consisting of bosons with mass $m_B$ and density $n_B$. 
 The distance between the layers is $d$ and the surface density of fermions 
in each layer is $n_\sigma$ with $\sigma=\uparrow,\downarrow$ denoting the two layers. When used in equations, they are taken to mean $\uparrow = +1, \downarrow = -1$, respectively. Occasionally we will also use the notation $\bs$ to mean the opposite layer from $\s$.
The boson-fermion interaction is short range and 
characterised by the strength $g=2\pi a_\text{eff}/\sqrt{m_rm_B}$ with $m_r=mm_B/(m+m_B)$ the reduced mass, and $a_\text{eff}$ the effective 2D-3D scattering 
length~\cite{Nishida2008}. Throughout, this interaction strength is taken to be weak in the sense $k_{F\sigma}a_\text{eff}\ll 1$, where $k_{F\sigma}=\sqrt{4\pi n_\sigma}$ is 
the Fermi momentum of layer $\sigma$. 

We treat the BEC using zero temperature Bogoliubov theory. This is a good approximation since the critical temperature of the superfluid phases 
of the fermions is much smaller than the critical temperature of the BEC. The bosonic degrees of freedom can then be integrated out, which yields an effective 
interaction between the fermions. In the static limit, this interaction is on the Yukawa form, and we end up with an effective Hamiltonian for the fermions in the two layers on 
the form~\cite{Heiselberg2000,Wu2016,Midtgaard2017}
\begin{align}
H= &H_\text{kin}+H_\text{int}=\sum_{\k, \s}  \xi_{\k \s} c^\d_{\k\s} c_{\k \s}\nn\\
&  +\frac{1}{2\mathcal{V}}\sum_{\sigma,\sigma'}\sum_{\k,\k' ,\q} V_{\sigma\sigma'}(\q) c^\d_{\k+\q\sigma} c^\d_{\k'-\q\sigma'} c_{\k'\sigma'} c_{\k\sigma}.
\end{align}
Here, $c^\d_{\k\s}$ creates a fermion in layer $\s$ with 2D momentum $\k=(k_x,k_y)$,  the dispersion in each layer is $\xi_{\k\sigma} = k^2/2m -\mu_\sigma$ with $\mu_\sigma$
the chemical potentials, and $\mathcal{V}$ is the system volume. We define the average chemical potential $\mu=(\mu_\uparrow+\mu_\downarrow)/2$ and 
the  ``magnetic field" $h=(\mu_\downarrow-\mu_\uparrow)/2$, so that we can write $\xi_{\k\sigma}=k^2/2m -\mu+\sigma\cdot h$.
The Yukawa interaction is
\begin{align}
V_{\sigma\sigma'}(\q) = - \frac{2g^2 n_B m_B}{\sqrt{q^2 + 2/\xi_B^2}} e^{ -\sqrt{k^2 + 2/\xi_B^2}\cdot |\s-\s'|d/2},
\label{Interaction}
\end{align}
where $\xi_B=1/\sqrt{8\pi n_Ba_B}$ is the healing length of the BEC with $a_B$ the boson-boson scattering length. The healing length determines 
the range of the induced interaction as can be seen by Fourier transforming Eq.~\eqref{Interaction} by to real space giving 
$V(r)=g^2 n_B m_B\pi^{-1}\exp( -\sqrt{2}r/\xi_B)$ with $r$ the 3D distance between the fermions. It follows that the range of the interaction can be tuned by varying the 
density $n_B$ or the scattering length $a_B$ of the surrounding BEC, which turns out to be a key property for the following. We note that retardation effects can  
be neglected, when the speed of sound in the BEC is much larger than the Fermi velocity in the two planes~\cite{Wu2016}.

\section{Pairing and Green's functions}
The attractive  interaction given by Eq.\ \eqref{Interaction}  can lead to Cooper pairing within each layer (intralayer pairing), and between the two layers  (interlayer pairing).
We recently analysed the 
  competition  between intra- and interlayer pairing for the case of equal density in each layer, i.e.\
  for $n_\uparrow=n_\downarrow$~\cite{Midtgaard2017}.
For a layer distance $d$ large compared to the range 
$\xi_B$ of the interaction, we found that the ground state is characterised by intralayer $p$-wave pairing, whereas $s$-wave interlayer pairing is stable for smaller $d/\xi_B$. 
In addition, we identified a crossover phase for intermediate $d/\xi_B$ where both types of pairing co-exist.  
 
 In this paper, we focus on interlayer $s$-wave pairing corresponding to  $d/\xi_B\lesssim 1$. We shall investigate the case of a non-zero field $h$ giving rise to a 
 population difference between the two layers, and the possibility of FFLO interlayer pairing with 
  non-zero COM momentum. Such interlayer pairing with COM momentum $\Q$ is characterised by 
 the  anomalous averages  $\langle c_{\Q/2+\k\s} c_{\Q/2-\k,\bar\s}\rangle$, which leads us to define the corresponding pairing field 
 \begin{align}
 \Delta_{\s\bar\s}(\k,\Q)=\frac1{\mathcal V}\sum_{\k'}V(\k-\k')\langle c_{\Q/2+\k'\s} c_{\Q/2-\k',\bar\s}\rangle.
 \label{DeltaDef}
 \end{align}
 We have dropped the subscripts on the induced interaction $V(\q)$, as it here and in the following refers to the interlayer interaction only, i.e. 
 $V(\q)\equiv V_{\sigma\bar\sigma}(\q)$. The pairing field obeys the Fermi 
 anti-symmetry $\Delta_{\s\bar\s}(\k,\Q)=-\Delta_{\bar\s\s}(-\k,\Q)$. In real space, it is on the form
  \begin{align}
 \Delta_{\s\bar\s}(\mathbf{r}_1,\mathbf{r}_2)=&V(\mathbf{r}_1-\mathbf{r}_2)\langle\psi_\s(\mathbf{r}_1)\psi_{\bar\s}(\mathbf{r}_2)\rangle\nn\\
 =&
 \frac1{\mathcal V}\sum_{\Q,\k}\Delta_{\s\bar\s}(\k,\Q)e^{i\Q\cdot{\mathbf R}}e^{i\k\cdot{\mathbf r}},
 \label{DeltaRealSpace}
 \end{align}
 where ${\mathbf R}=({\mathbf r}_1+{\mathbf r}_2)/2$ and ${\mathbf r}={\mathbf r}_1-{\mathbf r}_2$ are the COM and relative coordinates respectively, and 
 $\psi_\s(\mathbf{r})=\sum_{\k}c_{\k\s}\exp(\k\cdot\mathbf{r})$ is the field operator for particles in layer $\s$. We note that the pairing field is not translationally 
 symmetric in the FFLO phase, i.e.\ $\Delta_{\s\bar\s}(\mathbf{r}_1,\mathbf{r}_2)\neq \Delta_{\s\bar\s}(\mathbf{r}_1-\mathbf{r}_2)$. 
 The interaction   becomes in the mean-field  BCS approximation  
 \begin{align}
 H_\text{int}^\text{MF}=\sum_{\Q,\k} \Delta_{\uparrow\downarrow}(\k,\Q)c^\d_{\Q/2-\k\downarrow} c^\d_{\Q/2+\k\uparrow}+\text{h.c.},
 \end{align}
 where we include only the pairing channel as we focus on the superfluid instability. The Hartree-Fock terms will in general lead to 
 small effects in the weak coupling regime, which mostly can be accounted for by a renormalisation of the chemical potentials $\mu_\s$. 
 
All results presented here can be obtained by a direct diagonalisation of the 
mean-field BCS Hamiltonian $H^\text{MF}=H_0+H_\text{int}^\text{MF}$ using a standard Bogoliubov transformation. We however use  Green's functions to analyse the 
FFLO states, since this formalism naturally allows us to go beyond mean-field theory to include effects such as retardation, if needed in the future. 
The  normal and anomalous Green's functions for the superfluid phases are defined in  the standard way as 
\begin{align}
G_{\s} (\k,\k',\tau) &= - \langle T_\tau c_{\k\s}(\tau) c_{\k'\s}^\d (0) \rangle \nn\\
F_{\s} (\k,\k',\tau) &=  - \langle T_\tau c_{\k\s}(\tau) c_{-\k',\bar\s} (0) \rangle \nn\\
F^\d_{\s} (\k,\k',\tau) &= - \langle T_\tau c^\d_{-\k,\s}(\tau) c_{\k'\bar\s}^\d (0) \rangle,
\label{GreensDef}
\end{align}
where $T_\tau$ denotes imaginary-time ordering. Using $H^\text{MF}=H_\text{kin}+H_\text{int}^\text{MF}$, the Gor'kov equations for these Green's functions 
are straightforwardly derived. They read
\begin{widetext}
\begin{align}
\left( i\omega_n - \xi_{\k\s}  \right) G_{\s} (\k,\k',\omega_n) &= \delta_{\k,\k'}-  \sum_\Q \Delta_{\s\bs}(\k- \Q/2,\Q) F^\d_{\bs} (\k-\Q,\k',i\omega_n)  
\label{eq:EOMG} \\
\left( i\omega_n - \xi_{\k\s}  \right) F_{\s} (\k,\k',\omega_n) &=   \sum_\Q \Delta_{\s\bs}(\k-\Q/2,\Q) G_{\s} (-\k', -\k+\Q, -i\omega_n) 
\label{eq:EOMF} \\
\left( i\omega_n + \xi_{-\k,\s}  \right) F^\d_{\s} (\k,\k',\omega_n) &=  \sum_\Q \Delta^\ast_{\s\bs}(-\k-\Q/2,\Q) G_{\bs}(\k+\Q, \k',  i\omega_n),
\label{eq:EOMFd}
\end{align}
\end{widetext}
where we have Fourier transformed to Matsubara frequency space with $\omega_n=(2n+1)\pi T$. The self-consistent gap equation is then from \eqref{DeltaDef} and \eqref{GreensDef}
\begin{gather}
\Delta_{\s\bar\s}(\k,\Q) = -\frac{T}{\mathcal{V}} \sum_{\k',n} V(\k-\k')\nn\\
\times F_{\s}(\Q/2+\k'/2, \k'-\Q/2, i\omega_n) e^{-i\omega_n 0_+}.
\label{eq:deltadef}
\end{gather}

\section{Fulde-Ferrel Superconductivity}
 The $\Q$ values for which $\Delta_{\s\bar\s}(\k,\Q)$ is non-zero determines the structure of the order parameter in the superfluid phase. It has 
 been shown that $\Delta_{\s\bar\s}(\mathbf{r}_1,\mathbf{r}_2)$  can form very complicated 2D structures 
 corresponding to $\Delta_{\s\bar\s}(\k,\Q)\neq 0$  for many $\Q$s
 in Eq.\ \eqref{DeltaRealSpace}~\cite{Shimahara1998,Mora2004}. 
  However, the FFLO phase exhibits a second order transition to the normal phase in 2D at an upper critical field $h_{c2}$~\cite{Shimahara1994,Burkhardt1994}, 
 and at this transition it is  sufficient to consider the case of $\Delta_{\s\bar\s}(\k,\Q)\neq 0$ only for a 
 single $\Q$ vector. The reason is that  any linear combination of the $\Delta_{\s\bar\s}(\k,\Q)$'s, which are unstable towards pairing, is degenerate at $h_{c2}$, since it is 
 only the non-linear part of the gap equation that determines the optimal  combination  that minimises the energy. 
 
 In the following, we therefore consider the case  $\Delta_{\s\bar\s}(\k,\Q)\neq 0$ only for  a single $\Q$. This will give the correct upper critical field $h_{c2}$ 
for the second order transition between the  FFLO and the normal phase. We also expect that it will give a fairly precise value for the lower critical 
field $h_{c1}$ determining the first order transition between the FFLO and the superfluid phase, since the energy difference between  the FFLO phases with various spatial 
structures is  small~\cite{Shimahara1998,Mora2004}. Our scheme recovers the  usual homogenous BCS pairing for 
 $\Q=0$,   and   it corresponds to a plane wave Fulde-Ferrel (FF) type of pairing when  $\Q\neq0$, as can be seen from Eq.\ \eqref{DeltaRealSpace}. 
 Since we only have one $\Q$ vector, we can simplify the notation for the gap as $\Delta_{\s\bar\s}(\k,\Q) \to \Delta_{\s\bar\s}(\k)$.
 The Gor'kov equations \eqref{eq:EOMG}-\eqref{eq:EOMFd} 
 are then easily solved giving 
 \begin{align}
G_{\s}(\k,\k'i\omega_n) &=\delta_{\k,\k'} \left( i\omega_n - \xi_{\k,\s} - \frac{|\Delta_{\s\bs}(\k-\Q/2)|^2}{i\omega_n + \xi_{-\k+\Q,\bs}}\right)^{-1}\nn\\
F_{\s}(\k,\k',i\omega_n) &= \frac{\Delta_{\s\bs}(\k-\Q/2)}{ i\omega_n - \xi_{\k\s}}G_{\s}(\k,\k'i\omega_n)
\label{Fsolution}
\end{align}
 Using Eq.\ \eqref{Fsolution} in Eq.\ \eqref{eq:deltadef}, we obtain  the gap equation 
\begin{align}
\Delta(\k) = -\int\!d^2\check k' V(\k-\k') \frac{\Delta(\k')}{2 E_{\k'}}[1-f^+_{\k'}-f^-_{\k'}],
\label{eq:gapeq}
\end{align}
where $E^{\pm}_{\k'} = E_{\k'} \pm ( h + \frac{\k'\cdot\Q}{2m} )$, 
with $E_{\k'} = [( \k^{'2}/2m + \Q^2/8m - \mu )^2 + |\Delta(\k')|^2 ]^{1/2}$, and the Fermi distribution function is $f^{\pm}_{\k}=[\exp(\beta E^{\pm}_{\k})+1]^{-1}$.
In Eq.\ \eqref{eq:gapeq} and the rest of this paper, we have further simplified the notation by defining  $\Delta_{\uparrow\downarrow}(\k) \to \Delta(\k)$
and $d^2\check k'=d^2 k'/(2\pi)^2$. 
We solve Eq.\ \eqref{eq:gapeq} along with the number equation
\begin{equation}
N=\sum_\k \left[1-\frac{\xi_\k + q^2/8m}{E_\k} \left(1-f^+_{\k}-f^-_{\k}\right) \right].
\label{eq:numbereq}
\end{equation}
Note that in order to compare with previous results in the literature, we keep the \emph{total} number of particles, $N=N_\uparrow+N_\downarrow$, fixed, 
and not the number of particles in each plane. 
To solve the gap equation \eqref{eq:gapeq}, we perform a partial wave expansion of the induced  
interaction $V(\k) = \sum_n V_n(k) \exp (in\phi_\k)$, where $\phi_\k$ is the azimuthal angle of $\k$. In the numerics, we  keep the two leading terms, $n=0,1$ corresponding to $s$-wave singlet and $p$-wave triplet pairing respectively.
When solving the above equations, we find three different phases: The homogeneous BCS phase with $\Q=0$, the FF state with $\Q\neq0$,
 and the normal phase with no superfluid pairing. To determine which of these phases is the ground state, we compare their  energy $E$. We have from 
 mean-field theory  
\begin{gather}
E-\mu_\uparrow N_\uparrow-\mu_\downarrow N_\downarrow = \sum_\k \xi_\k-\sum_{E_\k^->0}E_\k^-+\sum_{E_\k^+<0}E_\k^+ \nn \\
+\sum_\k \frac{|\Delta(\k)|^2}{2E_\k}(1-f^+_{\k}-f^-_{\k}).
\label{Energy}
\end{gather}
%The summations over $E_\k^-$ and $E_\k^+$ only include $\k$ vectors, where they are  positive/negative respectively. 

%\georg{  As a sidenote, it is easy to check that the formula for the Green's function does not close for two or more $Q$-vectors. In this case, an approximate solution must be used.}

 \section{Results for zero temperature}
\begin{figure}[htb]
\centering
\includegraphics[width=\columnwidth]{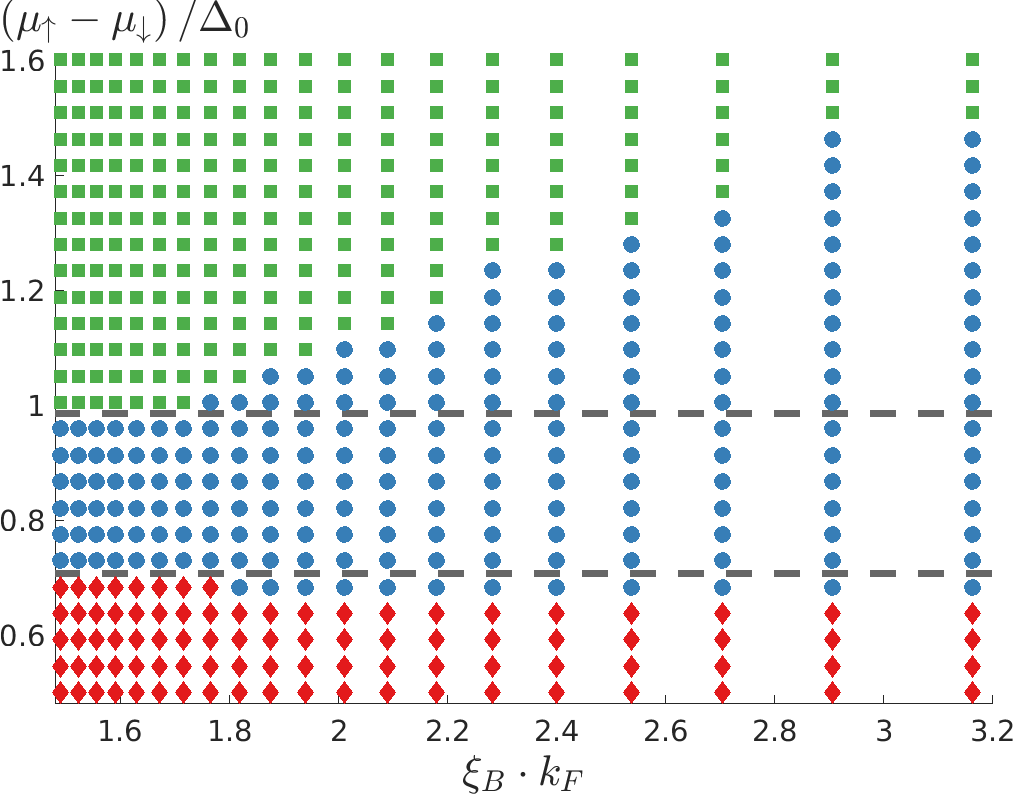}
\caption{(Color online) The $T=0$ phase diagram of the bilayer fermions for 
$k_Fd=1$ and $k_Fa_\text{eff}=0.05$,
 as a function of the interaction range $\xi_B$ and the field $h$.  The (red) diamonds, (blue) circles and (green) squares indicate the BCS, FF and normal phase, respectively. The horizontal dashed lines give the upper and lower critical fields for the FF phase for a short 
range interaction~\cite{Shimahara1994,Burkhardt1994}. }
\label{fig:xiBphases}
\end{figure}

\begin{figure}[htb]
\centering
\includegraphics[width=\columnwidth]{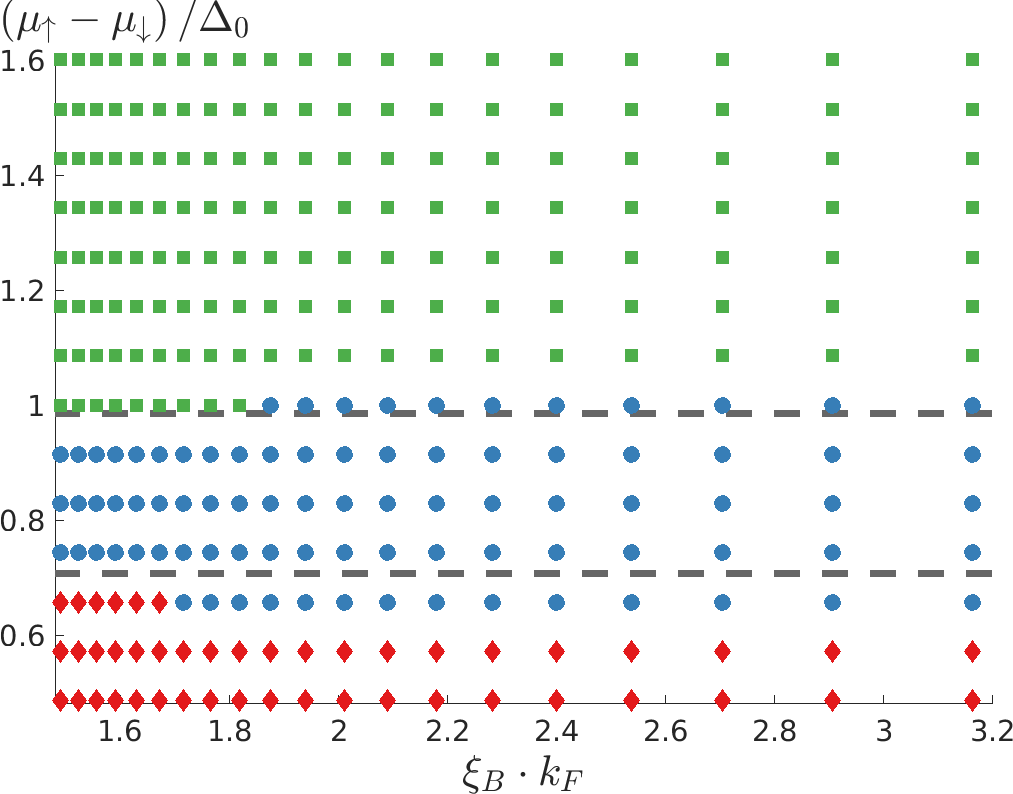}
\caption{(Color online) The $T=0$ phase diagram of the bilayer fermions as a function of the interaction range $\xi_B$ and the field $h$ for zero layer distance and boson-fermion interaction strength $k_Fa_\text{eff}=0.05$. The symbols and lines mean  
 the same as in Fig.~\ref{fig:xiBphases}.}
\label{fig:xiBphasesClose}
\end{figure}

 We now solve the Gor'kov equations  for $T=0$, varying the interaction range and density imbalance through 
 the healing length $\xi_B$ and the field $h = (\mu_\uparrow-\mu_\downarrow)/2$, respectively. The resulting phase diagrams are 
 shown in Figs.~\ref{fig:xiBphases} and~\ref{fig:xiBphasesClose}  for two different values of the layer distance, $k_Fd=1$ and 
 $k_Fd=0$.  The Fermi momentum is defined as $k_F = \sqrt{2\pi (n_\uparrow + n_\downarrow)}$. We set the effective scattering length to $k_Fa_\text{eff}=0.05$. This small value ensures that we stay in the valid range of mean-field theory for all parameter sets shown in the figures. As is standard in the literature, we measure the 
 field $h$ in units of $\Delta_0 = \Delta(k_F,\mathbf{0})$, which is the pairing field at the Fermi surface for $h=0$, that is, with no density imbalance.
 
Consider first the case of the layer distance $k_F d = 1$. Fig. \ref{fig:xiBphases} clearly shows  that there is a large region 
in the phase diagram, where the FF phase is stable. We find that the phase transition between the BCS and the 
FF phase  is first order at the lower critical field $h_{c_1}$, whereas it is second order for the transition between the FF phase and the normal phase at the upper critical 
field $h_{c_2}$. This is  in agreement with the results for a short range interaction~\cite{Shimahara1994,Burkhardt1994}. Moreover, 
the range of values of $h$ for which the FF phase is the ground 
state increases with the interaction range $\xi_B$. This shows that a long range interaction stabilises the FF phase. 
The reason is that the relative strength of the $p$-wave component  compared to the $s$-wave component of the interaction increases with increasing range, which favors FF pairing. To illustrate this important point further, we  plot as 
horizontal lines in Fig.~\ref{fig:xiBphases} the critical fields for a short range interaction~\cite{Fulde1964,Lee2017}, $h_{c_1} \approx \Delta_0/\sqrt{2} \approx 0.7 \Delta_0$ and $h_{c_2}= \Delta_0$. While the FF region approaches that 
of a short range interaction for decreasing $k_Fd$, it becomes much larger with increasing range $k_F\xi_B$.

Consider next the case of zero layer distance $d=0$ shown in Fig.~\ref{fig:xiBphasesClose}. While a large  $k_F\xi_B$ still stabilises the 
FF phase, the effect here is much less pronounced as compared to the case $k_Fd=1$. Increasing  $\xi_B$ leads to a smaller  
increase in the range of values of  $h$ for which the FF phase is the ground 
state  than for $k_Fd=1$. The reason is that the short range $1/r$ divergence of the Yukawa interaction between the fermions 
in the two layers given by Eq.~\eqref{Interaction}
 is cut off at $1/d$ for a finite layer distance $d$. This 
 makes the $p$-wave part of the interaction stronger compared to the $s$-wave part. As a result, the FF phase where there is pairing in both the $s$- and 
$p$-wave channels is favored compared to the pure $s$-wave BCS state for a non-zero layer distance. Note that the reason the superfluid region (BCS and FF) seems larger for 
$k_Fd=1$ compared to $d=0$ even though the strength of the interaction obviously  is smaller for a non-zero layer distance,
 is that we measure $h$ in units of $\Delta_0$, which is also
smaller. Had we used the unit $\epsilon_F$ instead for instance, the superfluid region of the $d=0$ phase would be larger.

 \section{Berezinskii-Kosterlitz-Thouless Melting}
\begin{figure}[htb]
\centering
\includegraphics[width=\columnwidth]{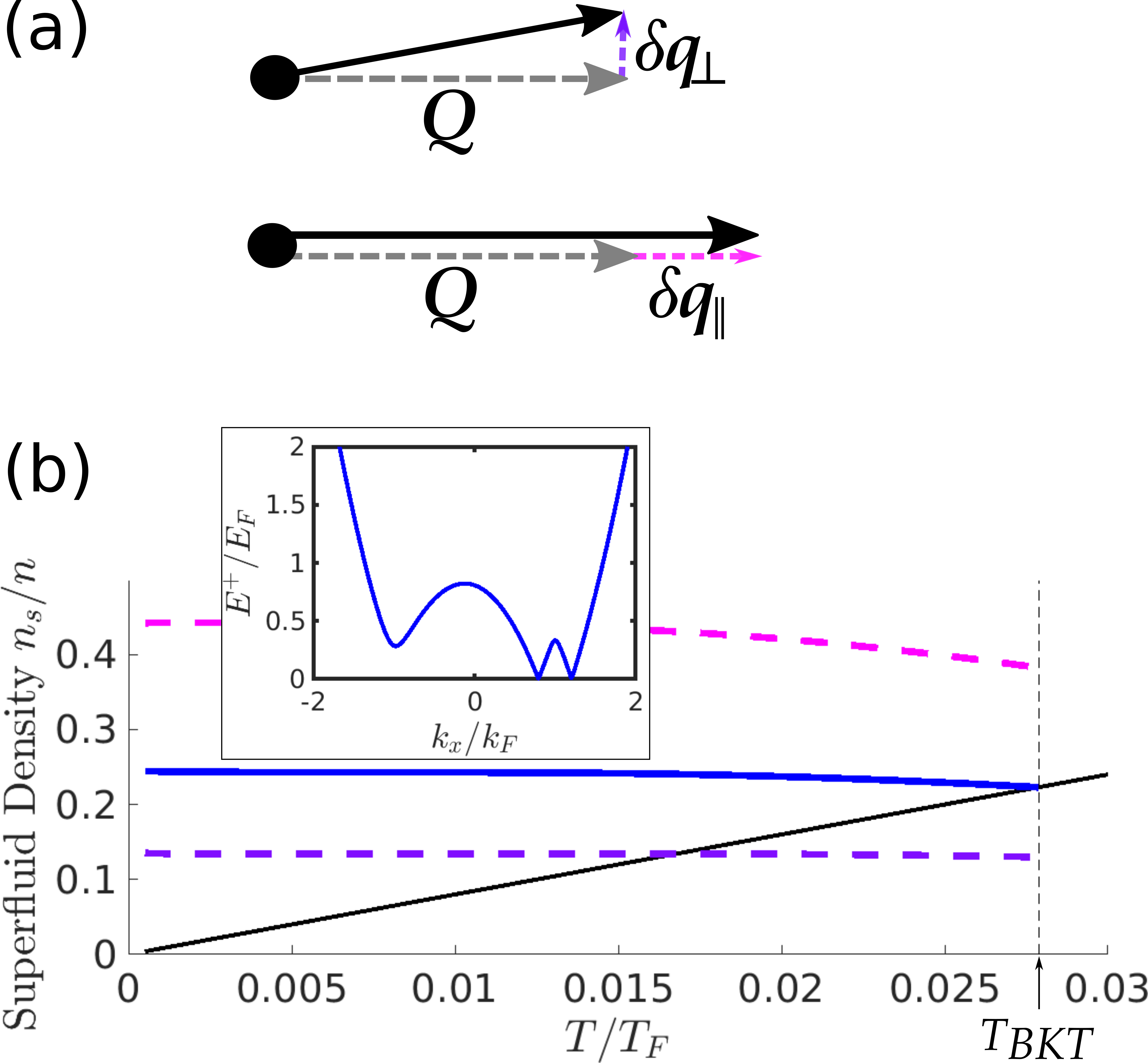}
\caption{(Color online) (a) The two kinds of phase fluctuations with $\delta\q$ perpendicular and parallel to the COM momentum $\Q$
of the Cooper pairs give rise to a rotation and a compression/expansion of the plane wave pairing field respectively. 
 (b) The corresponding superfluid densities $n_{s\parallel}$ (top dashed magenta line) and $n_{s\perp}$ (bottom dashed purple line). The solid blue line is  to the effective superfluid density $n_s=\sqrt{n_{s\parallel}n_{s\perp}}$. 
 The critical temperature $T_\text{BKT}$ is reached when $n_s$ crosses the thin solid  black line as indicated by the vertical dashed line. 
 The inset shows one branch of the quasiparticle spectrum along $k_y = 0$ with the COM momentum $\Q$ along the $x$-axis for $T=0$. The gapless excitations leading to the 
 reduction in the superfluid density are clearly visible.}
\label{fig:BKT}
\end{figure}

Since the fermions are confined in 2D layers, phase fluctuations of the order parameter are significant and will eventually melt the superfluid 
through the Berezinskii-Kosterlitz-Thouless (BKT) mechanism at a critical temperature $T_\text{BKT}$. To describe this, 
 we first calculate the superfluid stiffness, or equivalently the superfluid density, by 
imposing a linear phase twist on the order parameter and calculating the corresponding energy cost to  second order in the twist. 
For a given vector $\Q$, the real space pairing  becomes using Eq.\ \eqref{DeltaRealSpace}
\begin{align}
 \Delta(\mathbf{R},\mathbf{R})=\Delta \cdot e^{i(\Q+\delta\q)\cdot{\mathbf R}},
 \label{DeltaPhaseTwist}
 \end{align}
where $\Delta=\sum_{\k}\Delta(\k)/{\mathcal V}$ and $\delta\q\cdot{\mathbf R}=\delta\theta({\mathbf R})$ is the imposed spatially linear 
 phase twist. From Eq.\ (\ref{DeltaPhaseTwist}), it is clear that the direction  of the phase twist relative to the COM of the Cooper pairs is important: 
  When $\delta\q$ is parallel to $\Q$, the phase twist corresponds to adding/removing COM momentum to the Cooper pairs which compresses/expands 
the wavelength of the plane wave pairing field $ \Delta(\mathbf{R},\mathbf{R})$; when $\delta\q$ is perpendicular to $\Q$, the phase twist corresponds to a small rotation of the COM momentum to the Cooper pairs which rotates 
the plane wave pairing field. These two effects are illustrated in Fig.~\ref{fig:BKT}(a).
%%%%%%%%%%%%%%%%%%%%%%%%%%%%%%%%%%%%%%%%%%%%%%%%%%%%%%%%%%
%%%%%%%%%%%%%%%%%%%%%%%%%%%%%%%%%%%%%%%%%%%%%%%%%%%%%%%%%%

The phase twist $\delta\theta({\mathbf R})$  gives a free energy cost $\delta F$ of the form 
\begin{align}
\delta F=&\frac12\int\! d^2{\mathbf r}[J_\parallel(\partial_\parallel \delta\theta)^2+J_\perp(\partial_\perp \delta\theta)^2]\nn\\
=&\frac J2\int\! d^2{\mathbf r}(\nabla \delta\theta)^2,
\label{FreeEn}
\end{align}
where $\partial_\parallel$ and $\partial_\perp$ are  spatial derivatives parallel or perpendicular to the COM momentum of the Cooper pairs, corresponding to 
$\delta\q\parallel\Q$ and $\delta\q\perp\Q$ respectively. 
The associated superfluid stiffness constants are 
$J_\parallel$ and $J_\perp$. In  the second line of Eq.~\eqref{FreeEn}, we have rescaled the spatial coordinate perpendicular to $\Q$ by the factor 
$\sqrt{J_\parallel/J_\perp}$ to obtain an isotropic $XY$-model with the effective stiffness constant $J=\sqrt{J_\parallel J_\perp}$~\cite{Wu2016b}. 
Alternatively, defining the superfluid densities parallel and perpendicular to $\Q$ as 
$n_{s\perp}=4mJ_\perp$ and $n_{s\parallel}=4mJ_\parallel$, we can write the free energy cost as  
\begin{align}
\delta F=m\int\! d^2{\mathbf r}[n_{s\parallel}v_{s\parallel}^2+n_{s\perp}v_{s\perp}^2]/2.
\end{align} 
Here $v_{s\parallel}=\partial_\parallel\delta\theta/2m$ is the superfluid velocity parallel to $\Q$ and likewise for $v_{s\perp}$.
A long but straightforward calculation of the second order energy shift due to the phase twist gives
\begin{align}
n_{s\parallel} = n- \frac{\beta}{m} \int\!d ^2 \check k [f_{\k}^+ (1-f_{\k}^+)+f_{\k}^- (1-f_{\k}^-)] k_\parallel^2 
\label{ns}
\end{align} 
for the superfluid density along $\Q$, where $k_\parallel=\k\cdot\Q/Q$. Here, $n$ is the total surface density of fermions coming from the two layers. 
An equivalent formula holds for $n_{s\perp}$.  Equation \eqref{ns} is the 2D version of 
the usual expression for the superfluid density allowing  for the effects of the spatial anisotropy of the FF state~\cite{Leggett1975,Landau1980statistical}.
We can now determine the critical temperature of the superfluid using the BKT condition 
\begin{align}
T_\text{BKT}=\frac\pi2J=\frac\pi{8m}n_s
\label{eq:BKTcondition}
\end{align}
where we have defined the effective superfluid density as $n_s=\sqrt{n_{s\parallel}n_{s\perp}}$.

In Fig.~\ref{fig:BKT}(b), we plot the superfluid densities 
 as a function of $T$ for layer distance $k_Fd=1$ and boson-fermion interaction strength 
 $k_Fa_\text{eff}=0.05$. We see that $n_{s\parallel}>n_{s\perp}$, reflecting that the energy cost related to compressing/expanding  the 
 COM momentum is higher than that related to rotating it as expected. 
 Note that both superfluid densities are smaller than the total density $n$ even for $T\rightarrow 0$. 
This is due to the inherent presence of gapless quasiparticle states in the FF phase. These gapless excitations, which are 
shown in the inset of Fig.~\ref{fig:BKT}(b), reduce the superfluid density. 
We also plot the effective  superfluid density $n_s=\sqrt{n_{s\parallel}n_{s\perp}}$ as well
as the line $8mT/\pi$. It follows from Eq.\ (\ref{eq:BKTcondition}) that the superfluid phase melts when $n_s$ crosses this line.
\begin{figure}[htb]
\centering
\includegraphics[width=\columnwidth]{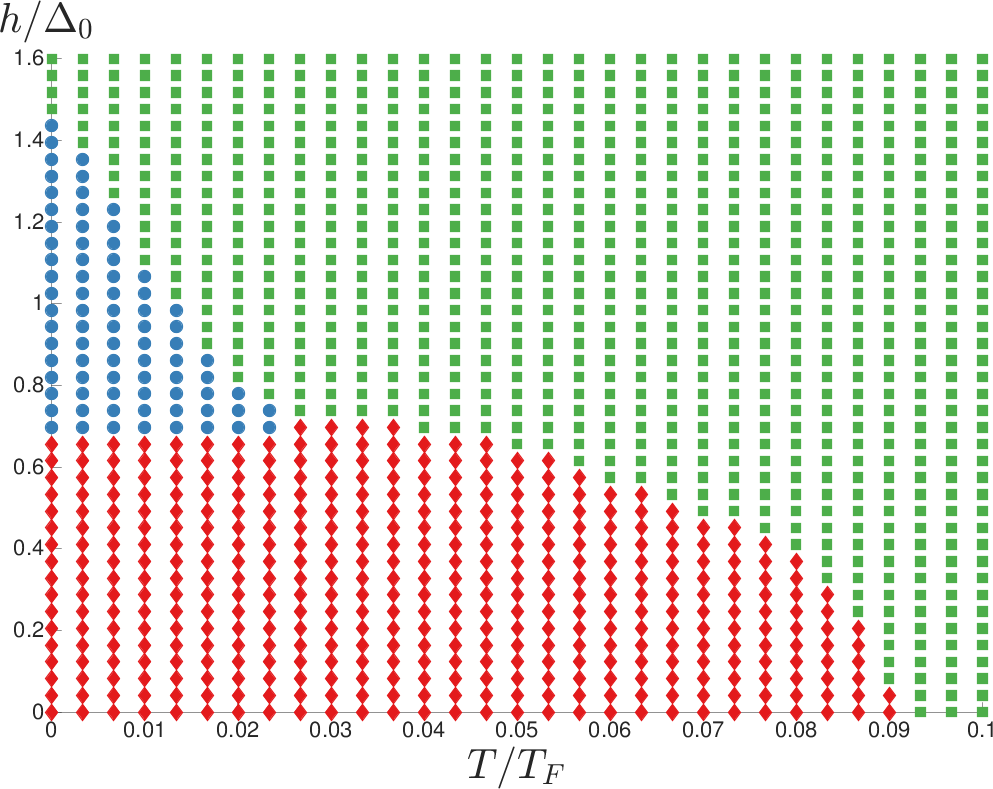}
\caption{(Color online) The  phase diagram as a function of temperature $T$ and field $h$ for layer distance $k_Fd=1$, 
boson-fermion interaction strength $k_Fa_\text{eff}=0.05$, and range $k_F\xi_B=3$. The symbols mean the same as 
in Figs.~\ref{fig:xiBphases}-\ref{fig:xiBphasesClose}.}
\label{fig:tempPhases}
\end{figure}

Having set up the theory for BKT melting, we can now analyse the BCS and FF phases at non-zero temperatures. 
In Fig.~\ref{fig:tempPhases}, we present an example of a phase diagram for $k_Fd=1$ and $k_Fa_\text{eff}=0.05$. We see that the critical 
temperature of the FF phase increases with decreasing field $h$. The highest critical temperature is obtained just above the lowest critical field
$h_{c_1}$, where the FF, BCS, and normal phase meet in a \emph{tricritical point} at $h_{c_1} \approx 0.7\Delta_0$ and $T\approx 0.025 T_F$.
This critical temperature is well below the theoretical maximum of $T_F/8$ obtained  by setting $n_s=n$ in  Eq.~(\ref{eq:BKTcondition}).
The reason is that the gapless excitations in the FF phase decrease the superfluid density below $n$, as we discussed 
above in connection with  Fig.~\ref{fig:BKT}. 
Indeed, we note that the value of the two components of the superfluid density in Fig.~\ref{fig:BKT} at $T=T_c$ are almost unchanged from their value at $T \rightarrow 0$.  
This is a general result:  While the flexibility of the bilayer system allows us to optimize the induced  interaction to favour 
 FF pairing, the gapless nature of this state prevents us from reaching critical temperatures close to $T_F/8$.

The critical temperature of the BCS phase is  much higher as can be seen from  Fig.~\ref{fig:tempPhases}. Due to its fully gapped spectrum, 
 $T_\text{BKT}$ can in fact relatively easily be tuned to be close to maximum $T_F/8$ by varying the layer distance $d$, 
 the interaction range $\xi_B$, even while keeping the boson-fermion interaction 
 strength $a_\text{eff}$ weak.

\section{Conclusion}
We analysed a mixed-dimensional system consisting of two layers of identical fermions immersed in a BEC. This system was shown to support superfluid 
pairing due to an attractive induced interaction between the two layers mediated by the BEC. When the densities of the two layers are different, the resulting superfluid 
phase is inhomogeneous. Using a plane wave FF ansatz to describe this phase, we demonstrated that it is stabilised by the nonzero range of the induced interaction. 
Importantly, the FF phase can occupy much larger regions of the phase diagram as compared to the  case of a short range interaction.  
The range of the induced interaction can be tuned by varying the BEC healing length, which makes the present system   promising for realising FFLO 
physics experimentally. We furthermore analysed the 
melting of the superfluid phases due to phase fluctuations using BKT theory, and demonstrated that the normal, homogenous and inhomogeneous
superfluid phases meet in a tricritical point in the phase diagram. The superfluid density of the FF phase was shown to be suppressed by  intrinsic gapless 
excitations, and this  leads to a significant reduction in the critical temperature compared to the homogeneous superfluid, which can be tuned to have a critical 
temperature close to the maximum $T_F/8$.

\acknowledgements
We wish to acknowledge the support of the Danish Council of Independent Research -- Natural Sciences via Grant No. DFF - 4002-00336.

\bibliography{Ref_fflo}

\end{document}